# DL based  analysis of movie reviews

Mary Pa, Amin Kazemi

Department of Electrical and Computer Engineering, Lakehead University, Canada

Department of mechanical and industrial engineering university of Toronto

Email: mpaparim@lakeheadu.ca, amin.kazemi@utoronto.ca

**Abstract**

Undoubtedly, social media are brainstormed by a tremendous volume of stories, feedback, reviews, and reactions expressed in various languages and idioms, even though some are factually incorrect. These motifs make assessing such data challenging, time-consuming, and vulnerable to misinterpretation. This paper describes a classification model for movie reviews founded on deep learning approaches. Almost 500KB pairs of balanced data from the IMDb movie review databases are employed to train the model. People's perspectives regarding movies were classified using both the long short-term memory (LSTM) and convolutional neural network (CNN) strategies. According to the findings, the CNN algorithm's prediction accuracy rate would be almost 97.4%. Furthermore, the model trained by LSTM resulted in accuracies of around and applying 99.2% within the Keras library. The model is investigated more by modification of model parameters. According to the outcomes, LTSM outperforms CNN in assessing IMDb movie reviews and is computationally less costly than LSTM.

## I-Introduction

Human communications are predominately based on languages. A commonly used way to share ideas is the most extensive worldwide communication network of computers, the internet. Users tend to present their thoughts, feelings, and emotions in formal and informal ways, resulting in a vast vocabulary domain, morphological diversity, and orthographic discrepancy. Due to the mentioned intrinsic features of the languages, it is challenging to process and analyze them reliably. Consequently, several recent attempts have been made to establish facilities for natural language processing (NLP) approaches. Sentiment recognition is a typical NLP response in several applications, including news categorization [1], spam email screening [2], and product review evaluation [3], which is widely used to fulfill the abovementioned aim. On the other hand, text classification is growing continually as one of the other most attractive applications of NLP due to vast unorganized text data in the web environment [4]. The procedure has a well-organized flow of work, starting with data collection and preprocessing, such as scaling, normalization, cleaning, and feature selection. Eventually, modelling, cost function, and employing good optimization algorithms are required to reach the goal. Several tools may be used to assess and interpret the trained model, such as a confusion matrix or many other scoring functions.

Convolutional neural networks (CNNs), recurrent neural networks (RNNs), and long short-term memory (LSTM) are commonly used solutions to investigate such a problem, apart from NLP [6]. The LSTM is an RNN framework that has been updated to alleviate the issue of short-term memory caused by the vanishing or exploding of gradients.Moreover, RNNs operate in a sequential pattern, computing the output, considering the previous stage and carrying the data over multiple computing steps. Regarding the issues mentioned earlier, RNNs are not capable enough to handle large sequences and data dependencies. The capacity of LSTM to keep and sustain long-term memory and the capability to employ context data from the previous step sequentially is hence its strength. Although, compared to the RNN, the LSTM structure is a bit more complicated.

Data scientists utilize CNN in natural language processing applications, such as sentence classification tasks [8]. It works the same way as a feature extractor, which stores semantic aspects of phrases. One of CNN's major drawbacks is the difficulty of determining the appropriate kernel window size.Some crucial data may be lost if adjusted too low. In contrast, the training process gets problematic if chosen to be high.

Numerous data scientists and software engineers across the globe are fascinated by the sentiment analysis of the IMDb review problem. The categorization of IMDb reviews has been the subject of several pieces of research so far. In contrast, the majority are unofficially reported. Reference [11] categorized the IMDb reviews and employed CNN and LSTM techniques and pre-trained word vectors from IMDb movies. The CNN classifier achieved 87% classification accuracy using two convolutional layers and two pooling layers, while the LSTM gave 81.8 % accuracy. A superior result of 88% was obtained by combining the abovementioned procedures. Reference [12], on the other hand, came to the opposite effect, claiming that the CNN algorithm surpassed both the LSTM and the combined LSTM-CNN approaches. Another research used a combination of CNN and bidirectional LSTM to boost accuracy to around 90%. According to [13], a support vector machine employing a tagged lexicon feature accurately classified the IMDb reviews by 87.1%. It was also concluded adopting bigrams and trigrams when processing the data boosted the accuracy by 2.1% more. In [14], the LSTM method obtained an accuracy rate of 89.9%, and the corresponding enhancement was due to accurate input data preprocessing. Using a random forest approach for the same classifying problem, [15] achieved an accuracy of 88.95%. In [16], an LSTM model applying a novel activation function was suggested to train the algorithm. Almost 93.5% of accuracy resulted from such a model. Although, to improve more, [17] obtained even better results by considering the decision tree algorithm to get about 94% accurate classification. In evaluating the IMDb movie ratings, an elevated level of accuracy was also reached, applying document embedding for training with cosine similarity. In this regard, reached 97.42% of the accurate solution.

In this article, a binary classifier was proposed to be used for sentiment analysis and to classify raw input text as positive or negative using two deep learning-based algorithms: LSTM and CNN. In particular, the IMDb dataset was utilized and freely accessible on different platforms such as GitHub. The contextual features are extracted using deep learning algorithms, and their polarity detection is examined. The proposed approach outperformed other research results mentioned in the literature and in different sections. The remaining of this research paper compromised the methodology section and then modelling the examination afterwards; experiments and conclusion are the final steps of this work.

## II-Methodology

In this section, different steps required to perform sentiment analysis are presented. The investigation was done by Google Colab using a personal computer with a processor Intel(R), Core(TM) i5-10400 CPU @ 2.90GHz, RAM 8.00 GB (7.79 GB usable) and the device ID is 00325-81966-19302-AAOEM. The system type is a 64-bit operating and has an x64-based processor.

### A: Data understanding

In this research, the IMDb dataset, which contains 500KB reviews, is used in such use cases as text analysis or sentiment analysis. The data set has already been classified into two groups, "positive" and "negative" reviews. Exploratory data analysis (EDA) is one of the most important aspects of data science. EDA is an approach to analyzing data sets to understand data that helps identify attributes, summarize their key characteristics, and discover patterns and spot anomalies in the dataset. In this work, data preparation is shown in Figure 1. In the data preparation stage, the goal is to clean the data and transform them into the required format accepted by the LSTM and CNN models.

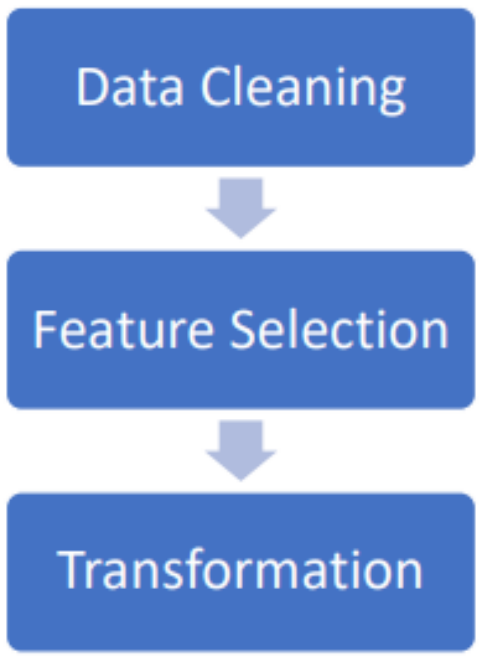

Figure 1 Sequence of data preparation steps in the deep learning network

Data cleaning is crucial in data preprocessing and identifies erroneous, duplicate, incomplete, or outdated data. Data cleaning is also applied to correct or remove inaccurate and corrupt entries, which can adversely affect the results. In the first step, the records were released with non-word characters (i.e., everything except numbers and letters). These include the symbols such as "?", "!", and "*". Subsequently, all runs of whitespaces were replaced with no space and digits with no slack. Furthermore, the reviews were tokenized to filter out the useless words, such as the stop words of the English language (i.e., articles, prepositions, pronouns, and conjunctions). The list of stop words that should be removed was imported from the natural language toolkit library (NLTK). The stop word removal helps eliminate the noise among the records and reduces the processing time. Finally, the model was run with and without removing the HTML tags and URLs. It was noticed that the prediction accuracy was negatively affected when they were removed (validation accuracy dropped to 50%). Moreover, the effects of lemmatization and stemming were tested, and the prediction accuracy decreased by a few percent.

**B: Feature selection and data transformation**

With the current dataset, only the columns "review" and "sentiment" will be used in this stage. To fit the data to the functionalities of our classifiers, LSTM and CNN, several techniques have been applied to transform data into the appropriate format required to be fed to the models; splitting the data into training, validation and testing folds, building vocabulary, padding, truncating, batching and loading as tensors and creating required layers. Before tokenizing and creating the vocabulary set, the dataset has spitted into training (first 70%) and testing (last 30%). It has been fixed so that our experiments are not affected by the randomness of the selected data.

The first step was tokenizing the sentences of the training and testing sets to build the vocab dictionary. The tokenize function takes all the train and the test sentences. It returns a NumPy array of the final training/testing list, a NumPy array of the training/testing labels, and the vocabulary set called" onehot_dict." The "onehot_dict" contains 1000 words. After tokenization, the length of the reviews was visualized to get the description summary of the variable. The minimum measurement size was 0, and the maximum length was 1000.

Calculating the length was helpful for the next step, which is padding. In fact, in recurrent neural networks such as RNN and LSTM, all the sequences should have the same length using any other framework. For this reason, they need to equalize their lengths by adding zeros to the end of the short ones. Note that there are other options to perform padding, such as right or left padding. After padding, created tensors from the "x_train" padded and the "y_train" using the "Tensor Dataset" function imported from the utils since the inputs should be 3D tensors in the framework.

Finally, the last step in the feature engineering is the creation of the data loaders, "train loader" and "valid loader," with a batch size of 1. The LSTM and CNN (with Keras and implementation) share the same feature engineering.

## III-Modelling

The parameters deployed in this LSTM are shown below. It includes an embedding layer and an LSTM layer. Furthermore, it contains a dropout layer as an effective layer to prevent overfitting and as a regularization technique. A ReLU layer was added to reshape the outputs, a dense layer and fully connected layer had the same roles. This architecture was applied during the function created. The parameters of the LSTM network are summarized in Table 1.

Table 1 The LSTM parameters

| **Layer** | **Parameters** |
|---|---|
| 1 | Input |
| 2 | Embedding |
| 3 | LSTM |
| 4 | FC |
| 5 | Activation |
| 6 | Dropout |
| 7 | Activation |
| 8 | Output |

The CNN sequential model was designed using the Keras framework: an embedding layer, 128,64 and 32 filters with the method same as padding, max-pooling dense layers, and activation function sigmoid with adaptive moment estimation (Adam) optimizer.

Table 2 contains the structure.

Table 2 The CNN parameters

| Layer | Parameters |
|---|---|
| 1 | Input |
| 2 | (Embedding) |
| 3 | (Conv1D),max pooling |
| 4 | (Conv1D),max pooling |
| 5 | (Conv1D),max pooling |
| 6 | (Flatten) |
| 7 | dense |
| 8 | (Dense) |

The CNN was applied on top of GLoVe[2] in the implementation process, a pre-trained word vectors model. First, the text data were tokenized to build the vocabulary dictionary named "word_id_dict" and to encode the sentences after padding them in the form of a NumPy array. This vocabulary The glossary helps make the embedding layer included in architecture after loading the pre-trained vectors for each token. After creating the Dataloaders using the DataLoader class to boost the training speed and save memory, the CNN model parameters were built, containing two filter sizes: pooling layers, 2 ReLU activation functions, and a dense layer for the flattening. This architecture was applied during the forward path using the on function created. After creating the embedding layer, the sequence is presented as a tensor that takes the token's ids as input. After that, the feature extractors of 1-dimensional were applied, yielding the feature maps as the output. The implementation of the abovementioned steps is demonstrated below.

The ReLU activation function and operation (Max pooling) is applied to reduce the architecture's computational complexity and cost and record the sequence's strongest signal. The resultant vector is then fed to the fully connected layer to decide which class is predicted. To minimize loss for both LSTM and CNN and to compute gradients, backpropagation was employed as usual for the deep learning models; the optimizer was the Adam op for both models. The auto grad was used to define both forward and backpropagation paths. At each epoch, a forward step was performed to compute the loss, followed by a backward step to add the gradients and update the weights/parameters of the model. A detailed explanation of metrics, including training accuracy, validation accuracy, and loss, is utilized in the code to keep track of the model's performance.

**V-Examination**

Further evaluation of the model was accomplished by creating a section for single-sentence predictions. First, the dataset was split based on the classes predicted into two data frames; (i) a data frame containing all positively predicted sentences and (ii) a data frame containing all negative sentences. After creating the two sub-data frames, we created a word cloud for each of them to display the most occurring positive and negative words. Then we applied a simple function that allows the user to select a phrase from the dataset and determine its predicted class. A close view of the most common words in the training and testing datasets is shown in Figure 2. As expected, the words "movie" and "film" are the most common words in both sets.

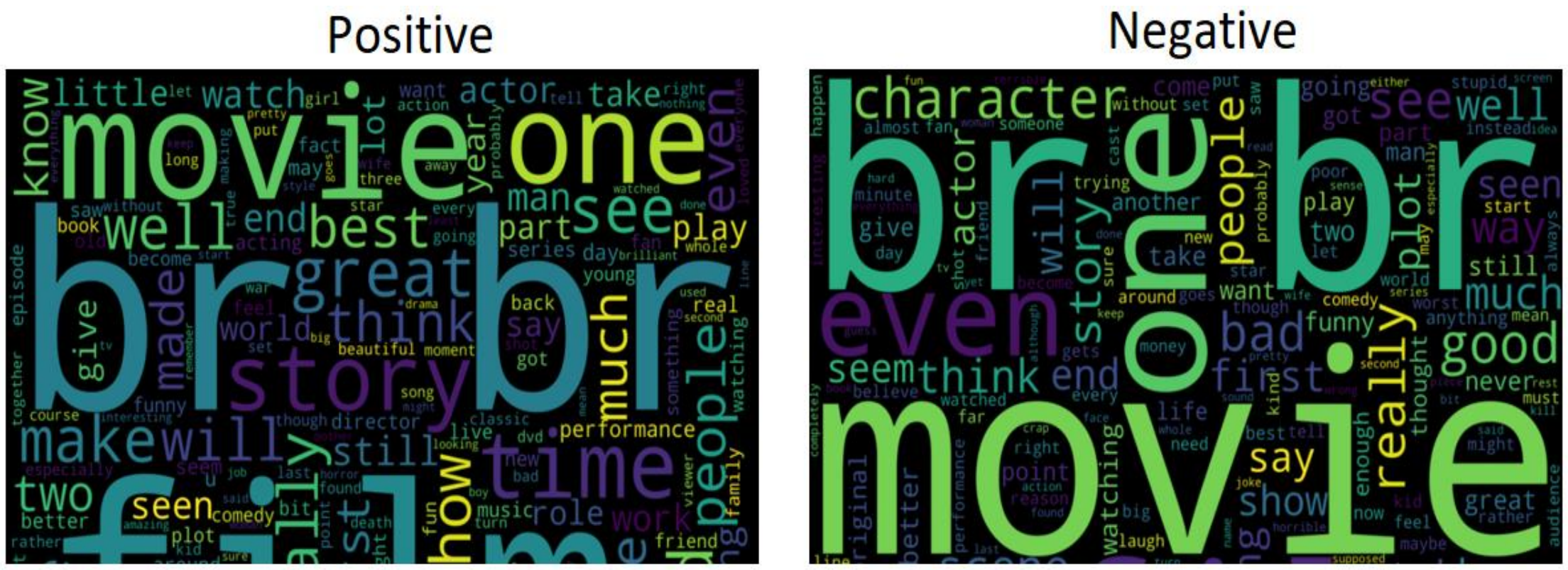

Figure 2 Most common words appeared in the positive and negative reviews in the tested data

The selected comment is as follows:

"*What an absolutely stunning movie; if you have 2.5 hrs to kill, watch it, you won't regret it, it's too much fun! Rajnikanth carries the movie on his shoulders, and although there isn't anything more other than him, I still like it. The music by A. R. Rehman takes time to grow on you, but after you heard it a few times, you really start liking it.*"

The actual sentiment for this statement is positive. It is found that the model sentiment is also positive, with a probability of 97.4%. After implementing our LSTM and CNN Keras architectures, have the results printed in detail and plotted using the Matplotlib library. A comparative study was performed to evaluate the prediction performance across the models and examined the version of each model by applying several changes. As a result, the validation accuracy of the LSTM model was 99.2%; meanwhile, the test accuracy was 97.6%On the hand, the validation accuracy of the CNN model was 97.4% within a 90.2% test accuracy.

### A:Effect of activation function

The impact of the activation function on the model predictions was explored. For this purpose, three different activation functions were used. As can be seen in the performances of the tanh and the sigmoid activation functions were almost the same, with the "ReLU" beings better than others.

### B:Effect of the initial learning rate

The effect of the initial learning rate was examined by setting it to 0.1 and 0.001. It can be concluded that the results would be better with the learning rate of 0.001 and setting the parameters explained earlier constant**.**

### C:Effect of the mini-batch size

The effect of the number of mini-batch sizes on the model performance is shown below. It is common to use 256 in these networks. The number was changed from 1 to 100 in the CNN LSTM network while keeping everything else constant. It was found that a network with 32 mini-batch size offers the best accuracy.

## VI-Conclusion

The LSTM and CNN classify the movie reviews from the IMDb dataset as positive or negative in this study. Different techniques of feature extraction that can capture the context of a word in a document with the semantic meanings and syntactic similarities and relation with other words were used with common preprocessing steps such as stop word and non-word removal and tokenization to improve the classification performance, followed by linguistic computation methods for the preprocessing of the data. It was found that lemmatization and stemming as well as the HTML tags and URL removal, did not improve the model accuracy. Therefore, they were excluded in the preprocessing step.

Although the applied methods provided satisfactory results, the models still have room to enhance the classification performance. Some more sophisticated algorithms, such as the combined LSTM with CNN, may give better results. Also, further steps in data cleaning, word embedding, pre-trained models, and other transfer learning approaches may provide better prediction accuracy and should be considered for future studies. Even though the movie reviews are quite ambiguous and do not necessarily follow grammatical rules, the results in the current study are encouraging in the context of natural language processing.

## Reference

[1] N.-T. Le, T. van Do, N. Thanh, N. Hoai, A. Le, and T. Editors, “Advances in Intelligent Systems and Computing 629 Advanced Computational Methods for Knowledge Engineering”, Accessed: Oct. 08, 2022. [Online]. Available: http://www.springer.com/series/11156

[2] M. H. Arif, J. Li, M. Iqbal, and K. Liu, "Sentiment analysis and spam detection in short informal text using learning classifier systems," Soft Comput., vol. 22, no. 21, pp. 7281–7291, Nov. 2018, doi: 10.1007/s00500-017-2729-x.

[3] P. Ray and A. Chakrabarti, "Twitter sentiment analysis for product review using lexicon method," in 2017 International Conference on Data Management, Analytics and Innovation (ICDMAI), Feb. 2017, pp. 211–216, doi: 10.1109/ICDMAI.2017.8073512.

[4] X. Guo, H. Zhang, H. Yang, L. Xu, and Z. Ye, "A Single Attention-Based Combination of CNN and RNN for Relation Classification," IEEE Access, vol. 7, pp. 12467–12475, 2019, doi: 10.1109/ACCESS.2019.2891770.

[5] S. Hochreiter and J. Schmidhuber, “Long Short-Term Memory,” Neural Comput., vol. 9, no. 8, pp. 1735–1780, Nov. 1997, doi: 10.1162/neco.1997.9.8.1735.

[6] A. Rehmer and A. Kroll, "On the vanishing and exploding gradient problem in Gated Recurrent Units," IFAC-PapersOnLine, vol. 53, no. 2, pp. 1243–1248, 2020, doi: 10.1016/j.ifacol.2020.12.1342.

[7] W. Yin, K. Kann, M. Yu, and H. Schütze, “Comparative Study of CNN and RNN for Natural Language Processing,” Feb. 2017, [Online]. Available: http://arxiv.org/abs/1702.01923.

[8] A. Hassan and A. Mahmood, "Deep Learning approach for sentiment analysis of short texts," in 2017 3rd International Conference on Control, Automation and Robotics (ICCAR), Apr. 2017, pp. 705–710, doi: 10.1109/ICCAR.2017.7942788.

[9] B. Pang and L. Lee, "A Sentimental Education: Sentiment Analysis Using Subjectivity Summarization Based on Minimum Cuts," Sep. 2004, [Online]. Available: http://arxiv.org/abs/cs/0409058.

[10] S. M. Qaisar, "Sentiment Analysis of IMDb Movie Reviews Using Long Short-Term Memory," in 2020 2nd International Conference on Computer and Information Sciences (ICCIS), Oct. 2020, pp. 1–4, doi: 10.1109/ICCIS49240.2020.9257657.

[11] V. K. and S. K., "Towards activation function search for long short-term model network: A differential evolution based approach," J. King Saud Univ. - Comput. Inf. Sci., May 2020, doi: 10.1016/j.jksuci.2020.04.015.

[12] T. P. Sahu and S. Ahuja, "Sentiment analysis of movie reviews: A study on feature selection & classification algorithms," in 2016 International Conference on Microelectronics, Computing and Communications (MicroCom), Jan. 2016, pp. 1–6, doi: 10.1109/MicroCom.2016.7522583.

[13] T. Thongtan and T. Phienthrakul, "Sentiment Classification Using Document Embeddings Trained with Cosine Similarity," in Proceedings of the 57th Annual Meeting of the Association for Computational Linguistics: Student Research Workshop, 2019, pp. 407–414, doi: 10.18653/v1/P19-2057.